\documentclass[aps,prl,twocolumn,amsmath,amssymb,superscriptaddress,longbibliography,nofootinbib]{revtex4-2}

\usepackage[T1]{fontenc}
\usepackage[latin9]{inputenc}
\usepackage{color}
\usepackage{verbatim}
\usepackage{lmodern}
\usepackage{xcolor}
\usepackage{amsmath}
\usepackage{amssymb}
\usepackage{graphicx}
\usepackage{nicematrix}
\usepackage{bm}
\usepackage{mathtools}
\usepackage{tikz}
\usepackage{diagbox}
\usepackage[normalem]{ulem}

\NiceMatrixOptions{cell-space-limits=1pt}

\usetikzlibrary{
  arrows.meta,
  positioning,
  calc,
  fit,
  backgrounds,
  decorations.pathmorphing,
  decorations.pathreplacing,
  shapes.geometric
}

\definecolor{copyA}{HTML}{0072B2}
\definecolor{copyB}{HTML}{E69F00}
\definecolor{measurecol}{HTML}{D55E00}
\definecolor{residualcol}{HTML}{CC79A7}
\definecolor{spectatorcol}{HTML}{777777}

\makeatletter
\PassOptionsToPackage{caption=false}{subfig} 
\usepackage{hyperref}
\hypersetup{
breaklinks=true,
colorlinks=true,
citecolor=blue,
linkcolor=black,
filecolor=black,
urlcolor=blue
}
\IfFileExists{lmodern.sty}{\usepackage{lmodern}}{}

\renewcommand{\hat}{\widehat}
\renewcommand{\leq}{\leqslant}

\newcommand{\CC}{\mathbb{C}}

\newtheorem{theorem}{Theorem}

\makeatletter

\newcommand*{\wideboxed}[1]{\setlength{\fboxsep}{1ex}%
  \fbox{\m@th$\displaystyle#1$}}
\makeatother

\def\be{\begin{equation}}
\def\ee{\end{equation}}

\def\c{\chi}

\makeatother

\begin{document}

\title{Exact Entanglement Swapping through Single-Occupancy Measurements in Gaussian Fermion States}

\author{Jiyuan Fang}
\affiliation{School of Physics, Georgia Institute of Technology, Atlanta, GA 30332, USA}


\begin{abstract}

We determine the exact entanglement structure of the conditional state obtained by measuring $m$ corresponding rungs ($m\leq N/2$) in two identical copies of an arbitrary half-filled free fermion Gaussian state and post-selecting the same normalized single-fermion state ($|\psi\rangle = u|10\rangle+v|01\rangle$, here 0 and 1 denote the fermion occupancy on each sites) on each rung. For $m<N/2$, the conditional wavefunction generally depends on the initial state. Nevertheless, whenever the selected outcome has nonzero probability, the state on the unmeasured sites remains Gaussian and factorizes exactly into $N-m$ different orthogonal modes including $m$ inter-copy entangling modes and $N-2m$ spectator modes localized in one copy. Consequently, the entanglement entropy between the unmeasured parts of the two copies is $S=m h_2(|v|^2)$, independent of the initial state, where $h_2(x)=-x\ln x-(1-x)\ln(1-x)$. The success probability is given by $P_m=\det C_L\det(I_m-C_L)=\det(C_{LR}C_{RL})$, determined solely by the initial correlations and independent of ($u$,$v$). Equal-weight Bell post-selection serves as a special case that achieves the maximal entanglement swapping.

\end{abstract}
\maketitle

\section{Introduction}\label{Sec:Introduction}

Quantum measurement is not merely a readout of a preexisting state, it is a nonunitary operation that can reorganize correlations already distributed through a composite system. Entanglement swapping is the clearest operational example: A Bell measurement on intermediate degrees of freedom can entangle systems that have never interacted directly, converting correlations initially stored in separate pairs into nonlocal entanglement between the residual systems \cite{Zukowski1993,Pan1998,Bose1998,Huhtanen2026}. This naturally raises a broad many-body question: how can measurements create, redistribute, and reshape entanglement in an extended quantum state?

Several lines of work address this question from complementary perspectives. Studies of localizable entanglement \cite{Verstraete2004,Popp2005} and single-round measurements \cite{Garratt2023,Weinstein2023,Sahay2025,KhannaVasseur2026} have shown how local measurements can concentrate correlations or prepare nontrivial many-body states \cite{Lu2022,Tantivasadakarn2024}. Entanglement swapping and teleportation have been extended to critical many-body systems \cite{Hoshino2025,Sala2024,Cheng2024}, while monitored systems reveal measurement-induced changes in entanglement dynamics and phase structure \cite{Li2018,Skinner2019,Cao2019,Alberton2021,Poboiko2023,Leung2025}. These studies mainly ask how much entanglement measurements can generate, how it scales, or how it evolves under repeated measurements. A more microscopic question is less understood: for one specified measurement outcome, what does the resulting many-body wavefunction look like, and which of its features, if any, are independent of the initial state?

Fermionic Gaussian states provide an analytically tractable setting in which states and entanglement can be described directly at the mode level \cite{Bravyi2005,BoteroReznik2004,Peschel2003,PeschelEisler2009}. In our previous work, Ref.~\cite{fang2025universal}, we identified a concrete puzzle in this setting. Consider two identical half-filled free-fermion states and post-select the same single-occupancy outcome on corresponding sites (rungs) of the two copies. When doing Bell post-selection measurements on exactly half of the rungs, the post-measurement state can only take a single universal form that carries the maximal amount of entanglement. For fewer measured rungs (still selecting the same states), different initial Slater determinants generally produced different post-measurement wavefunctions, while numerical results showed the same linear entropy law across those states.

In this work, we resolve this puzzle by deriving the exact conditional state for $m$ measured rungs, with $m$ not exceeding half of the system, and showing that the otherwise different post-measurement wavefunctions share the same underlying entanglement structure. This structure directly explains the universal entropy law: each measured rung contributes the same amount of entanglement, determined solely by the selected single-occupancy state and independent of the initial Slater determinant. We further derive an exact formula for the probability of obtaining the selected outcome. Unlike the entanglement, this probability is determined entirely by the initial states and is independent of the selected single-occupancy state. Within this framework, equal-weight Bell post-selection provides a particularly important special case: it maximizes the conditional inter-copy entanglement without changing the branch probability, yielding $S_{A_R}=m\ln 2$ and thereby realizing maximal entanglement swapping.

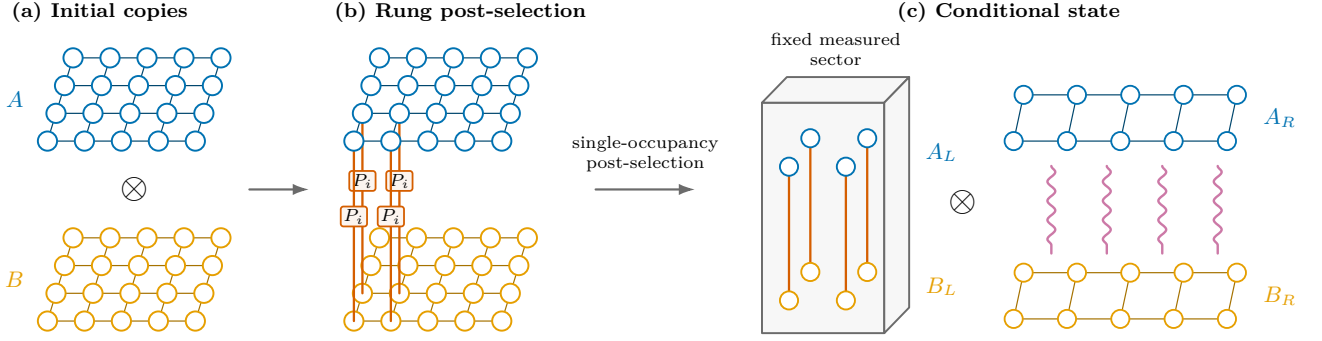
\begin{figure*}[htbp]
  \centering
  \resizebox{0.97\textwidth}{!}{%
    \begin{tikzpicture}[
  x=1cm,y=1cm,
  >=Latex,
  font=\footnotesize,
  line cap=round,
  line join=round,
  bondA/.style={draw=copyA!62!black,line width=0.48pt},
  bondB/.style={draw=copyB!72!black,line width=0.48pt},
  siteA/.style={circle,draw=copyA,fill=white,line width=0.78pt,
                inner sep=0pt,minimum size=2.70mm},
  siteB/.style={circle,draw=copyB,fill=white,line width=0.78pt,
                inner sep=0pt,minimum size=2.70mm},
  projectorline/.style={draw=measurecol,line width=0.92pt},
  projector/.style={rectangle,rounded corners=1.1pt,
                    draw=measurecol,fill=measurecol!6,
                    line width=0.72pt,minimum width=3.7mm,
                    minimum height=2.8mm,inner sep=0.15pt},
  paneltitle/.style={font=\bfseries\footnotesize,anchor=west},
  stagearrow/.style={-{Latex[length=2.30mm,width=1.60mm]},
                     draw=black!58,line width=0.88pt},
  cuboid/.style={draw=black!58,line width=0.78pt,fill=black!3},
  fixedrung/.style={draw=measurecol,line width=0.92pt},
  entangle/.style={draw=residualcol,line width=0.96pt,
                   decorate,decoration={snake,amplitude=0.55mm,
                                        segment length=2.7mm}}
]

\newcommand{\largegrid}[4]{%
  \begin{scope}[shift={(#1,#2)}]
    \foreach \r in {0,...,3}{
      \foreach \c in {0,...,3}{
        \draw[#3]
          ({0.52*\c+0.12*\r},{0.39*\r}) --
          ({0.52*(\c+1)+0.12*\r},{0.39*\r});
      }
    }
    \foreach \r in {0,...,2}{
      \foreach \c in {0,...,4}{
        \draw[#3]
          ({0.52*\c+0.12*\r},{0.39*\r}) --
          ({0.52*\c+0.12*(\r+1)},{0.39*(\r+1)});
      }
    }
    \foreach \r in {0,...,3}{
      \foreach \c in {0,...,4}{
        \node[#4] at ({0.52*\c+0.12*\r},{0.39*\r}) {};
      }
    }
  \end{scope}%
}

\newcommand{\residualgrid}[4]{%
  \begin{scope}[shift={(#1,#2)}]
    \foreach \r in {0,1}{
      \foreach \c in {0,...,3}{
        \draw[#3]
          ({0.75*\c+0.14*\r},{0.65*\r}) --
          ({0.75*(\c+1)+0.14*\r},{0.65*\r});
      }
    }
    \foreach \c in {0,...,4}{
      \draw[#3] ({0.75*\c},{0}) -- ({0.75*\c+0.14},{0.65});
    }
    \foreach \r in {0,1}{
      \foreach \c in {0,...,4}{
        \node[#4,minimum size=2.55mm,line width=0.74pt]
          at ({0.75*\c+0.14*\r},{0.65*\r}) {};
      }
    }
  \end{scope}%
}

\node[paneltitle] at (0.12,5.60) {(a) Initial copies};
\largegrid{0.74}{3.78}{bondA}{siteA}
\largegrid{0.74}{1.25}{bondB}{siteB}
\node[font=\bfseries,copyA,anchor=east] at (0.52,4.365) {$A$};
\node[font=\bfseries,copyB,anchor=east] at (0.52,1.835) {$B$};
\coordinate (A-layer-center) at (1.96,4.365);
\coordinate (B-layer-center) at (1.96,1.835);
\node[font=\Large] at ($ (A-layer-center)!0.5!(B-layer-center) $) {$\otimes$};
\draw[stagearrow] (3.58,3.10) -- (4.42,3.10);

\node[paneltitle] at (4.66,5.60) {(b) Rung post-selection};

\foreach \c/\r in {0/0,1/0,0/1,1/1}{
  \coordinate (Atop-\c-\r) at
    ({5.05+0.52*\c+0.12*\r},{3.78+0.39*\r});
  \coordinate (Bbot-\c-\r) at
    ({5.05+0.52*\c+0.12*\r},{1.25+0.39*\r});
}

\largegrid{5.05}{1.25}{bondB}{siteB}

\foreach \c/\r/\py in {0/1/3.22,1/1/3.22}{
  \draw[projectorline] (Bbot-\c-\r) -- (Atop-\c-\r);
  \node[projector,font=\scriptsize]
    at ({5.05+0.52*\c+0.12*\r},\py) {$P_i$};
}

\foreach \c/\r/\py in {0/0/2.72,1/0/2.72}{
  \draw[projectorline] (Bbot-\c-\r) -- (Atop-\c-\r);
  \node[projector,font=\scriptsize]
    at ({5.05+0.52*\c+0.12*\r},\py) {$P_i$};
}

\largegrid{5.05}{3.78}{bondA}{siteA}

\draw[stagearrow] (8.45,3.10) -- (9.85,3.10);
\node[font=\scriptsize,align=center,anchor=south] at (9.15,3.24)
  {single-occupancy\\post-selection};

\node[paneltitle,anchor=center] at (14.26,5.60) {(c) Conditional state};

\path[cuboid] (10.79,1.08) rectangle (12.51,4.32);
\path[cuboid] (10.79,4.32) -- (11.14,4.68) --
              (12.86,4.68) -- (12.51,4.32) -- cycle;
\path[cuboid] (12.51,1.08) -- (12.86,1.44) --
              (12.86,4.68) -- (12.51,4.32) -- cycle;
\node[font=\scriptsize,align=center] at (11.83,5.08)
  {fixed measured\\sector};

\foreach \xt/\yt/\xb/\yb in {
  11.17/3.42/11.17/1.54,
  11.97/3.42/11.97/1.54,
  11.47/3.82/11.47/1.94,
  12.27/3.82/12.27/1.94}{
  \draw[fixedrung] (\xb,\yb) -- (\xt,\yt);
  \node[circle,draw=copyA,fill=white,line width=0.76pt,
        minimum size=2.55mm,inner sep=0pt] at (\xt,\yt) {};
  \node[circle,draw=copyB,fill=white,line width=0.76pt,
        minimum size=2.55mm,inner sep=0pt] at (\xb,\yb) {};
}

\node[font=\bfseries,copyA,anchor=west] at (12.98,3.62) {$A_L$};
\node[font=\bfseries,copyB,anchor=west] at (12.98,1.74) {$B_L$};

\node[font=\Large] at (13.61,2.93) {$\otimes$};

\residualgrid{14.33}{3.78}{bondA}{siteA}
\residualgrid{14.33}{1.28}{bondB}{siteB}
\node[font=\bfseries,copyA,anchor=west] at (17.73,4.105) {$A_R$};
\node[font=\bfseries,copyB,anchor=west] at (17.73,1.605) {$B_R$};

\foreach \xx in {14.86,15.64,16.42,17.20}{
  \draw[entangle] (\xx,3.43) -- (\xx,2.20);
}

\end{tikzpicture}%
  }
  \caption{Schematic setup and protocol. (a) Two identical, initially decoupled, particle-number-conserving Gaussian states occupy copies $A$ and $B$; the two-dimensional lattices are schematic. (b) The same single-occupancy projector $P_i^{(u,v)}$ is post-selected on corresponding rungs. A contiguous $2\times2$ patch is shown only for visual clarity; the measured set need not have this geometry. (c) In a successful branch, the measured rungs form a fixed product sector, while the unmeasured copies $A_R$ and $B_R$ contain the residual intercopy entanglement. The intercopy connectors are schematic: the entangled residual modes are generally nonlocal orbitals within $R$, rather than site-by-site pairs. }
  
  \label{fig:setup}
\end{figure*}


\section{Setup and protocol}
\label{Sec:Setup}

We consider $N$ fermionic modes, with $N$ even. Let $c_i^\dagger$ and $c_i$ create and annihilate a fermion in mode $i$, respectively. They satisfy the canonical anticommutation relations 
\[
\{c_i,c_j^\dagger\}=\delta_{ij},\qquad
\{c_i,c_j\}=\{c_i^\dagger,c_j^\dagger\}=0,
\]and $c_i|\mathrm{vac}\rangle=0$. The operators $\gamma_\mu^\dagger$ create the occupied single-particle orbitals of the Slater determinant and are linear combinations of the site-mode creation operators $c_i^\dagger$. A particle-number-conserving pure fermionic Gaussian state is a Slater determinant of $k$ occupied single-particle orbitals.
At half filling, $k=N/2$, we write
\begin{equation}
|\Psi_0\rangle
=
\prod_{\mu=1}^{k}\gamma_\mu^\dagger|{\rm vac}\rangle,
\qquad
\gamma_\mu^\dagger
=
\sum_{i=1}^{N}c_i^\dagger (U_0)_{i\mu},
\qquad
k=\frac{N}{2}.
\label{Eq:InitialState}
\end{equation}
Here the $\mu$-th column of $U_0\in\mathbb C^{N\times k}$ is the wavefunction of the $\mu$-th occupied orbital. We choose these orbitals orthonormal, $U_0^\dagger U_0=I_k$, so that $\{\gamma_\mu,\gamma_\nu^\dagger\}=\delta_{\mu\nu}$.

We define the single-particle correlation matrix by
\[
C_{ij}
=
\langle\Psi_0|c_j^\dagger c_i|\Psi_0\rangle,
\]
for which $C=U_0U_0^\dagger$.

More generally, the occupied orbitals may be changed by any invertible matrix $G\in GL(k,\mathbb C)$:
\begin{equation}
\widetilde U=U_0G.
\end{equation}
Equivalently, the transformed occupied-orbital operators are
\[
\widetilde\gamma_\mu^\dagger
=
\sum_{\nu=1}^{k}\gamma_\nu^\dagger G_{\nu\mu}.
\]
Fermionic antisymmetry then gives
\[
\prod_{\mu=1}^{k}\widetilde\gamma_\mu^\dagger
|\mathrm{vac}\rangle
=
\det(G)
\prod_{\mu=1}^{k}\gamma_\mu^\dagger
|\mathrm{vac}\rangle.
\]
Thus $\widetilde U$ changes only the basis chosen within the
occupied single-particle subspace and represents the same normalized Slater determinant. For a nonorthonormal representative $\widetilde U$, the correlation matrix is
\begin{equation}
C=\widetilde U
(\widetilde U^\dagger\widetilde U)^{-1}
\widetilde U^\dagger .
\end{equation}
We use this occupied-column freedom only in the derivations in the Appendices.

In our measurement setup, we prepare two initially uncorrelated copies of this state, labeled $A$ and $B$,
\[
|\Psi_{\mathrm{in}}\rangle
=
|\Psi_0\rangle_A\otimes|\Psi_0\rangle_B.
\]
The operators $c_{i,A}^\dagger$ and $c_{i,B}^\dagger$ create a fermion in mode $i$ of copies $A$ and $B$, respectively. The pair of corresponding modes $(i,A)$ and $(i,B)$ will be called rung $i$. After relabeling the modes, the measured set is
\begin{equation}
L=\{1,\ldots,m\},
\qquad
0\leq m\leq k,
\label{Eq:MeasuredRegion}
\end{equation}
and $R=L^c$ is the unmeasured set. We denote the modes in copy $A$ belonging to $L$ and $R$ by $A_L$ and $A_R$, respectively, and similarly define $B_L$ and $B_R$.

For an single copy orbital $\chi=\sum_{x\in R}\chi_x e_x$ ($e_x$ is the site-orbital basis vector) supported on $R$,  we define the corresponding creation operator in copy $X=A,B$ by
\[
\chi_X^\dagger
=
\sum_{x\in R}\chi_x c_{x,X}^\dagger.
\]
On rung $i$, we use the occupation basis
\[
|n_A n_B\rangle_i
=
(c_{i,A}^\dagger)^{n_A}
(c_{i,B}^\dagger)^{n_B}
|\mathrm{vac}\rangle_i,
\qquad n_A,n_B\in\{0,1\},
\]
with copy $A$ ordered before copy $B$.
Thus
\[
|10\rangle_i=c_{i,A}^\dagger|\mathrm{vac}\rangle_i,
\qquad
|01\rangle_i=c_{i,B}^\dagger|\mathrm{vac}\rangle_i.
\]
The single-occupancy subspace is the sector
$n_{i,A}+n_{i,B}=1$.
\begin{equation}
|\phi(u,v)\rangle_i
=
u|10\rangle_i+v|01\rangle_i,
\qquad
|u|^2+|v|^2=1,
\label{Eq:SingleOccupancyState}
\end{equation}
The same coefficients $u$ and $v$ are used on every measured rung; this is what we mean by uniform post-selection. The corresponding occupied mode is
\begin{equation}
w_i^\dagger(u,v)
=
u c_{i,A}^\dagger+v c_{i,B}^\dagger,
\qquad
|\phi(u,v)\rangle_i=w_i^\dagger(u,v)|{\rm vac}\rangle_i.
\label{Eq:MeasuredModeGeneral}
\end{equation}
Within the single-occupancy sector, the two equal-weight choices are the following Bell states: 
\begin{equation}
\begin{aligned}
|{\rm Bell}_s\rangle_i
&=
\frac{|01\rangle_i+s|10\rangle_i}{\sqrt2}
=w_{s,i}^\dagger|{\rm vac}\rangle_i,\\
w_{s,i}^\dagger
&=
\frac{c_{i,B}^\dagger+s c_{i,A}^\dagger}{\sqrt2}.
\end{aligned}
\label{Eq:BellStateDefinition}
\end{equation}
with $s=\pm1$, corresponding to $(u,v)=(s,1)/\sqrt2$ up to an overall phase.

Operationally, each measured rung is measured in a basis containing $|\phi(u,v)\rangle_i$, and we retain only runs in which this outcome is obtained on every $i\in L$.

The single-rung and many-rung projectors are
\begin{equation}
P_i^{(u,v)}
=
|\phi(u,v)\rangle_i\langle\phi(u,v)|,
\qquad
\Pi_L^{(u,v)}
=
\prod_{i\in L}P_i^{(u,v)}.
\label{Eq:PostselectionOperator}
\end{equation}
We denote the measured product state by $|\Phi_L(u,v)\rangle=\bigotimes_{i\in L}|\phi(u,v)\rangle_i$. The success probability is

\begin{equation}\label{Eq:SuccessProbabilityDefinition}
P_m=
\langle\Psi_{\rm in}|
\Pi_L^{(u,v)}
|\Psi_{\rm in}\rangle.
\end{equation}
Whenever $P_m>0$, the normalized conditional state is
\begin{equation}\label{Eq:NormalizedPostselectedState}
    |\Psi_{\rm post}^{(u,v)}\rangle
=
P_m^{-1/2}
\Pi_L^{(u,v)}
|\Psi_{\rm in}\rangle,
\qquad P_m>0.
\end{equation}
We call this normalized conditional state the successful
post-selected branch. Throughout, entanglement refers to the von Neumann entropy associated with the fermionic mode partition $A_R:B_R$.

\section{Exact conditional form and maximal swapping}
\label{Sec:CanonicalResult}

We now state the canonical form of every nonzero-probability conditional state.

\begin{theorem}[Canonical form]
\label{Thm:CanonicalForm}
Assume $P_m>0$.  Then there exist orthonormal orbitals
$\{\alpha_\ell\}_{\ell=1}^{k-m}$ and
$\{\beta_j\}_{j=1}^{m}$ supported on $R$ such that
\begin{equation}
\langle\alpha_\ell,\alpha_{\ell'}\rangle
=\delta_{\ell\ell'},
\quad
\langle\beta_i,\beta_j\rangle
=\delta_{ij},
\quad
\langle\alpha_\ell,\beta_j\rangle=0.
\label{Eq:CanonicalOrthonormality}
\end{equation}
Define the residual modes
\begin{equation}
h_j^\dagger(u,v)
=
u^*\beta_{j,B}^\dagger-v^*\beta_{j,A}^\dagger.
\label{Eq:CanonicalNonlocalMode}
\end{equation}
With a fixed fermionic ordering, the complete normalized post-selected state has the canonical form
\begin{align}
|\Psi_{\rm post}^{(u,v)}\rangle
\doteq{}&
\left(\prod_{i=1}^{m}w_i^\dagger(u,v)\right)
\left(\prod_{j=1}^{m}h_j^\dagger(u,v)\right)
\notag\\
&\times
\left(\prod_{\ell=1}^{k-m}
\alpha_{\ell,A}^\dagger\alpha_{\ell,B}^\dagger\right)
|{\rm vac}\rangle.
\label{Eq:CanonicalStateGeneral}
\end{align}
The three products in Eq.~\eqref{Eq:CanonicalStateGeneral} have distinct roles. The first fixes the $m$ measured rungs in the selected state. The second contains the m residual modes that carry all entanglement between $A_R$ and $B_R$. The third contains $k-m$ spectator orbitals, each occupied once in copy $A$ and once in copy $B$; 
Equivalently,
\begin{equation}
|\Psi_{\rm post}^{(u,v)}\rangle
\doteq
|\Phi_L(u,v)\rangle\otimes|\psi_R^{(u,v)}\rangle,
\label{Eq:CanonicalStateFactorization}
\end{equation}
where $|\psi_R^{(u,v)}\rangle$ is obtained from Eq.~\eqref{Eq:CanonicalStateGeneral} by omitting the measured-mode product.  Here $\doteq$ denotes equality up to the global phase fixed by the fermionic ordering. 
\end{theorem}

This result separates the unmeasured conditional state into two physically distinct sectors. The $\alpha^\dagger$ sector of orbitals occupied independently in both copies and therefore carries no intercopy entanglement. The $h^\dagger$ sector consists of $m$ orthogonal orbital pairs, each containing one fermion coherently shared between copies $A$ and $B$, which carry all the residual intercopy entanglement. We refer to these as the spectator and residual sectors, respectively. Those spatial orbitals $\alpha_\ell$ and $\beta_j$ generally depend on the initial Slater determinant.

\emph{Physical mechanism.--} The mechanism can be understood from a single measured rung. After restricting to the one-particle sector, there are two coherent possibilities. Either the fermion on the measured rung occupies copy $A$, while the associated residual orbital lies in $B_R$, or the measured fermion occupies copy $B$, while the residual orbital lies in $A_R$. Projecting the local rung onto $u|10\rangle+v|01\rangle$ fixes the measured mode and leaves a residual superposition\eqref{Eq:RawResidualModeMain}. 

This process can be summarized by this following formula:
\begin{equation}\label{Eq:SingleRungMechanismMain}
P_i^{(u,v)}
d_{i,A}^\dagger d_{i,B}^\dagger
|\mathrm{vac}\rangle
=
w_i^\dagger(u,v)\,
\eta_i^\dagger(u,v)
|\mathrm{vac}\rangle,
\end{equation}
with
\begin{equation}
\eta_i^\dagger(u,v)
=
u^*\widehat v_{i,B}^\dagger
-
v^*\widehat v_{i,A}^\dagger.
\label{Eq:RawResidualModeMain}
\end{equation}
As shown in detail in Appendix~\ref{App:CanonicalProof}, we apply an allowed change of occupied orbital basis that isolates one occupied-orbital pair for each measured rung, which makes this single-rung measurement picture possible. 

The construction extends independently to all measured rungs. In the adapted occupied-orbital basis of Appendix~\ref{App:CanonicalProof}, each measured rung is associated with one pair of occupied modes, one from each copy. The local projector maps this pair to one fixed measured mode and one residual mode supported on $R$, so iterating the projection preserves the Slater-determinant form.

The resulting residual orbitals are generally not orthogonal. However, uniformity is crucial: every residual mode has the same relative amplitudes $(-v^*,u^*)$ between copies $A$ and $B$. Their spatial wavefunctions can therefore be orthogonalized without changing these copy-space amplitudes. The remaining occupied orbitals form $k-m$ copy-separable spectator pairs. This produces the canonical state in Eq.~\eqref{Eq:CanonicalStateGeneral}.

The canonical form also makes the residual inter-copy entanglement transparent. Because the spectator modes are copy-separable, all entanglement between $A_R$ and $B_R$ resides in the $m$ mutually orthogonal inter-copy modes,  
\begin{equation}
h_j^\dagger(u,v)|{\rm vac}\rangle_{\beta_j}
=
u^*|01\rangle_{\beta_j}
-
v^*|10\rangle_{\beta_j}.
\label{Eq:ResidualTwoModeState}
\end{equation}
Here $|n_A n_B\rangle_{\beta_j}$ denotes the occupation basis of the two orbital modes $\beta_{j,A}$ and $\beta_{j,B}$, using the same $A$-before-$B$ convention. Let
\[
\rho_{A_R}
=
{\rm Tr}_{B_R}
|\psi_R^{(u,v)}\rangle
\langle\psi_R^{(u,v)}|,
\qquad
S_{A_R}
=
-{\rm Tr}\!\left(\rho_{A_R}\ln\rho_{A_R}\right).
\]
Because the $\beta_j$ orbitals are mutually orthogonal, those $m$ modes contribute additively, giving
\begin{equation}
\boxed{
S_{A_R}=m h_2(|v|^2).
}
\label{Eq:VonNeumannEntropyMain}
\end{equation}
Here $h_2(p)=-p\ln p-(1-p)\ln(1-p)$ is the binary entropy.  

\emph{Bell post-selection and maximality.--}
For Bell choice $(u,v)=(s,1)/\sqrt2,s=\pm1$, the projected single-occupancy state becomes one of the two Bell states in the single-occupancy sector. In this case, each residual mode may be chosen, up to an irrelevant column phase, as 
\begin{equation}
h_{s,j}^\dagger
=
\frac{
\beta_{j,B}^\dagger
-
s\beta_{j,A}^\dagger
}{\sqrt{2}}.
\label{Eq:BellResidualMode}
\end{equation}
Thus each factor in Eq.~\eqref{Eq:ResidualTwoModeState} is the Bell state $(|01\rangle_{\beta_j}-s|10\rangle_{\beta_j})/\sqrt{2}$ of modes $\beta_{j,A}$ and $\beta_{j,B}$, with the sign opposite to that of the selected rung. Hence
\begin{equation}
\boxed{
S_{A_R}=m\ln 2.
}
\label{Eq:BellEntropy}
\end{equation}

This value is also the largest entropy attainable from an $m$-rung selected branch, independently of the Gaussian construction.  The product post-selection state $|\Phi_L(u,v)\rangle$ has Schmidt rank at most $2^m$ across $A_L:B_L$.  Since the initial doubled state is a product across copies $A:B$, contracting its measured modes with $\langle\Phi_L(u,v)|$ produces a residual state of Schmidt rank at most $2^m$ across $A_R:B_R$.  Therefore, every selected branch obeys
\begin{equation}
S_{A_R}
\leq
\ln(2^m)
=
m\ln 2.
\label{Eq:SchmidtRankBound}
\end{equation}
For equal-weight Bell post-selection , the canonical state has exactly $2^m$ equal nonzero Schmidt coefficients and therefore saturates this bound.  

\emph{Why uniformity matters.--}
Uniformity means that every measured rung is projected onto the same state $u|10\rangle+v|01\rangle$. Consequently, all raw residual modes carry the same relative amplitudes $(-v^*,u^*)$ between copies $A$ and $B$. Orthogonalizing their spatial wavefunctions therefore changes only the spatial orbitals and leaves these copy-space amplitudes unchanged. If different rungs are projected onto different states, the residual modes no longer share a common pair of copy-space amplitudes, and their orthogonalization generally mixes the spatial and copy structures.  The resulting entanglement generally depends on the initial state. Appendix~\ref{App:NonuniformOutcomes} illustrates this point for mixed Bell-sign outcomes: maximal entanglement survives only when the projected residual Gram matrix is block diagonal between the two sign sectors. The entanglement of a mixed-sign branch is therefore generally initial-state dependent and need not be maximal.

\section{Success probability and Special cases}
\label{Sec:Probability}

\emph{Success probability.--}

Let $C_L$ be the $m\times m$ principal submatrix of the
single-copy correlation matrix on the measured set,
\begin{equation}\label{Eq:CLDefination}
    (C_L)_{ij}=C_{ij},\qquad i,j\in L
\end{equation}

and let $I_m$ denote the $m\times m$ identity matrix. The normalization of the raw Slater determinant in Appendix~\ref{App:ProbabilityProof} gives the gauge-invariant result
\begin{equation}
\boxed{
P_m
=
\det C_L\,\det(I_m-C_L),}
\label{Eq:ProbabilityCorrelation}
\end{equation}
Let $\nu_1,\ldots,\nu_m$ are the eigenvalues of $C_L$, then
\begin{equation}
P_m
=
\prod_{j=1}^{m}\nu_j(1-\nu_j)
\leq
4^{-m},
\label{Eq:ProbabilityBound}
\end{equation}
with equality if and only if $C_L=I_m/2$.  For a pure Slater determinant, $C^2=C$.  With respect to the decomposition of the single-copy mode space into the measured and unmeasured regions, $L\oplus R$, we write
\begin{equation}
C=
\begin{pmatrix}
C_L & C_{LR}\\
C_{RL} & C_R
\end{pmatrix},
\end{equation}
the $L$ block of $C^2=C$ gives
\begin{equation}
C_L(I_m-C_L)=C_{LR}C_{RL}.
\label{Eq:CorrelationCrossBlockIdentity}
\end{equation}
Consequently,
\begin{equation}
P_m=\det(C_{LR}C_{RL}).
\label{Eq:ProbabilityCrossCorrelations}
\end{equation}
The success probability is therefore controlled by the singular values of the one-body correlation block connecting the measured and unmeasured regions of a single copy.  It is nonzero exactly when $C_{LR}$ has full row rank $m$; if a natural mode on $L$ is deterministically empty or filled, the associated cross-boundary singular value vanishes and the selected branch has zero probability. Let $U_L\in\mathbb C^{m\times k}$ and
$U_R\in\mathbb C^{(N-m)\times k}$ be the row submatrices of $U_0$
restricted to $L$ and $R$, respectively, so that $U_0=(U_L^{\mathsf T},U_R^{\mathsf T})^{\mathsf T}$, the equivalent orbital expression is
\begin{equation}
P_m
=
\det(U_L U_L^\dagger)
\det(U_R^\dagger U_R).
\label{Eq:ProbabilityOrbitalMatrix}
\end{equation}

\emph{Bell choice at fixed success probability.--}
One may observe that equation~\eqref{Eq:ProbabilityCorrelation} contains no dependence on $u$ or $v$. By contrast, Eq.~\eqref{Eq:VonNeumannEntropyMain} is maximized at $|u|=|v|=1/\sqrt2$ . Equal-weight post-selection therefore realizes the maximal value $m\ln2$ at exactly the same branch probability as any other normalized uniform single-occupancy post-selection.  The initial cross-boundary correlations determine the probability of the selected branch, whereas the pair $(u,v)$ fixes the Schmidt weights of the normalized residual factors.

\emph{Half-measurement case.--}
When $m=k=N/2$, the spectator sector in Eq.~\eqref{Eq:CanonicalStateGeneral} is absent and the orbitals $\{\beta_j\}_{j=1}^{k}$ form a complete orthonormal basis of $R$. Because all residual modes have the same copy-space coefficients, a unitary change from the $\beta$-orbital basis to the site basis of $R$ changes the many-body state only by a global phase. The post-selected state therefore takes the rung-factorized form
\begin{equation}\label{Eq:HalfMeasurementState}
|\Psi_{\rm post}\rangle
\doteq
\bigotimes_{i\in L}|\phi(u,v)\rangle_i
\otimes
\bigotimes_{j\in R}|\phi(-v^*,u^*)\rangle_j .
\end{equation}
Here
\begin{equation}
|\phi(-v^*,u^*)\rangle
=
-v^*|10\rangle+u^*|01\rangle
\end{equation}
is the normalized single-occupancy state orthogonal to $|\phi(u,v)\rangle$. Thus every measured rung is fixed in the selected state, and every unmeasured rung carries its orthogonal partner. 

For the Bell choice $(u,v)=(s,1)/\sqrt{2}$, Eq.~\eqref{Eq:HalfMeasurementState} reduces to

\begin{equation}
|\Psi_{\rm post}^{(s)}\rangle
\doteq
\bigotimes_{i\in L}|{\rm Bell}_s\rangle_i
\otimes
\bigotimes_{j\in R}|{\rm Bell}_{-s}\rangle_j.
\label{Eq:HalfMeasurementBellState}
\end{equation}
Since both $U_L$ and $U_R$ are $k\times k$ matrix, Eq.~\eqref{Eq:ProbabilityOrbitalMatrix} becomes
\begin{equation}
P_k
=
|\det U_L\,\det U_R|^2
\label{Eq:HalfMeasurementProbability}
\end{equation}
recovering the half-measurement result of Ref.~\cite{fang2025universal}.

 \section{Conclusion and Discussion}\label{Sec:Conclusion}

In this work, we have determined the exact structure of the conditional state generated by uniform single-occupancy post-selection on corresponding rungs of two identical copies of a half-filled, particle-number-conserving Slater determinant. Whenever the selected branch has nonzero probability, the conditional state remains Gaussian and admits an exact canonical form in the unmeasured sector. This structure renders the entanglement between the unmeasured copies additive, yielding an entropy that grows exactly linearly with the number of measured rungs. The success probability, by contrast, is independent of the normalized uniform target state and is fixed entirely by the initial one-body correlations across the measured--unmeasured cut. As a special case, equal-weight Bell post-selection attains maximal conditional entanglement without changing the branch probability, thereby realizing maximal entanglement swapping. The central physical message is: \textbf{The selected rung state determines how much intercopy entanglement a successful branch carries, whereas the initial one-body correlations determine how likely that branch is to occur.} 

A natural next step is to extend this framework beyond uniform post-selection by characterizing nonuniform outcome strings and the complete measurement-outcome ensemble. The present analysis shows that generic nonuniform Bell-outcome strings do not share the state-independent maximality of the uniform branch; however, it remains open whether such branches admit a broader outcome-dependent canonical structure and how the branch-averaged conditional entanglement scales with the number of measured rungs. From an experimental perspective, few-rung post-selection may offer a more accessible test of the theory than half-system post-selection, whose success probability is necessarily exponentially suppressed with system size.

\emph{Acknowledgments. --}I thank Xueda Wen for helpful discussions and valuable comments on the manuscript.

\bibliography{ref}

@article{fang2025universal,
  title={Universal and Maximal Entanglement Swapping in General Fermionic Gaussian States},
  author={Fang, Jiyuan and Tang, Qicheng and Wen, Xueda},
  journal={arXiv preprint arXiv:2512.15890},
  year={2025}
}

@article{Zukowski1993,
  author  = {{\.{Z}}ukowski, Marek and Zeilinger, Anton and Horne, Michael A. and Ekert, Artur K.},
  title   = {{``Event-Ready-Detectors''} {Bell} Experiment via Entanglement Swapping},
  journal = {Phys. Rev. Lett.},
  volume  = {71},
  number  = {26},
  pages   = {4287--4290},
  year    = {1993},
  doi     = {10.1103/PhysRevLett.71.4287},
  url     = {https://doi.org/10.1103/PhysRevLett.71.4287}
}

@article{Pan1998,
  author  = {Pan, Jian-Wei and Bouwmeester, Dik and Weinfurter, Harald and Zeilinger, Anton},
  title   = {Experimental Entanglement Swapping: Entangling Photons That Never Interacted},
  journal = {Phys. Rev. Lett.},
  volume  = {80},
  number  = {18},
  pages   = {3891--3894},
  year    = {1998},
  doi     = {10.1103/PhysRevLett.80.3891},
  url     = {https://doi.org/10.1103/PhysRevLett.80.3891}
}

@article{Verstraete2004,
  author  = {Verstraete, Frank and Popp, Markus and Cirac, J. Ignacio},
  title   = {Entanglement versus Correlations in Spin Systems},
  journal = {Phys. Rev. Lett.},
  volume  = {92},
  number  = {2},
  pages   = {027901},
  year    = {2004},
  doi     = {10.1103/PhysRevLett.92.027901},
  url     = {https://doi.org/10.1103/PhysRevLett.92.027901}
}

@article{Popp2005,
  author  = {Popp, Markus and Verstraete, Frank and Mart{\'i}n-Delgado, Miguel A. and Cirac, J. Ignacio},
  title   = {Localizable Entanglement},
  journal = {Phys. Rev. A},
  volume  = {71},
  number  = {4},
  pages   = {042306},
  year    = {2005},
  doi     = {10.1103/PhysRevA.71.042306},
  url     = {https://doi.org/10.1103/PhysRevA.71.042306},
  eprint  = {quant-ph/0411123},
  archivePrefix = {arXiv}
}

@article{Garratt2023,
  author  = {Garratt, Samuel J. and Weinstein, Zack and Altman, Ehud},
  title   = {Measurements Conspire Nonlocally to Restructure Critical Quantum States},
  journal = {Phys. Rev. X},
  volume  = {13},
  number  = {2},
  pages   = {021026},
  year    = {2023},
  doi     = {10.1103/PhysRevX.13.021026},
  url     = {https://doi.org/10.1103/PhysRevX.13.021026}
}

@article{Sahay2025,
  author  = {Sahay, Rahul and Verresen, Ruben},
  title   = {Classifying One-Dimensional Quantum States Prepared by a Single Round of Measurements},
  journal = {PRX Quantum},
  volume  = {6},
  number  = {1},
  pages   = {010329},
  year    = {2025},
  doi     = {10.1103/PRXQuantum.6.010329},
  url     = {https://doi.org/10.1103/PRXQuantum.6.010329},
  eprint  = {2404.16753},
  archivePrefix = {arXiv},
  primaryClass  = {quant-ph}
}

@article{Hoshino2025,
  author  = {Hoshino, Masahiro and Oshikawa, Masaki and Ashida, Yuto},
  title   = {Entanglement Swapping in Critical Quantum Spin Chains},
  journal = {Phys. Rev. B},
  volume  = {111},
  number  = {15},
  pages   = {155143},
  year    = {2025},
  doi     = {10.1103/PhysRevB.111.155143},
  url     = {https://doi.org/10.1103/PhysRevB.111.155143}
}

@article{Sala2024,
  author  = {Sala, Pablo and Murciano, Sara and Liu, Yue and Alicea, Jason},
  title   = {Quantum Criticality under Imperfect Teleportation},
  journal = {PRX Quantum},
  volume  = {5},
  number  = {3},
  pages   = {030307},
  year    = {2024},
  doi     = {10.1103/PRXQuantum.5.030307},
  url     = {https://doi.org/10.1103/PRXQuantum.5.030307}
}

@article{Skinner2019,
  author  = {Skinner, Brian and Ruhman, Jonathan and Nahum, Adam},
  title   = {Measurement-Induced Phase Transitions in the Dynamics of Entanglement},
  journal = {Phys. Rev. X},
  volume  = {9},
  number  = {3},
  pages   = {031009},
  year    = {2019},
  doi     = {10.1103/PhysRevX.9.031009},
  url     = {https://doi.org/10.1103/PhysRevX.9.031009},
  eprint  = {1808.05953},
  archivePrefix = {arXiv},
  primaryClass  = {cond-mat.stat-mech}
}

@article{Leung2025,
  author  = {Leung, Chun Y. and Meidan, Dganit and Romito, Alessandro},
  title   = {Theory of Free Fermions Dynamics under Partial Postselected Monitoring},
  journal = {Phys. Rev. X},
  volume  = {15},
  number  = {2},
  pages   = {021020},
  year    = {2025},
  doi     = {10.1103/PhysRevX.15.021020},
  url     = {https://doi.org/10.1103/PhysRevX.15.021020}
}

@article{Bravyi2005,
  author  = {Bravyi, Sergey},
  title   = {Lagrangian Representation for Fermionic Linear Optics},
  journal = {Quantum Inf. Comput.},
  volume  = {5},
  number  = {3},
  pages   = {216--238},
  year    = {2005},
  doi     = {10.26421/QIC5.3-3},
  url     = {https://doi.org/10.26421/QIC5.3-3},
  eprint  = {quant-ph/0404180},
  archivePrefix = {arXiv}
}

@article{BoteroReznik2004,
  author  = {Botero, Alonso and Reznik, Benni},
  title   = {{BCS}-like Modewise Entanglement of Fermion Gaussian States},
  journal = {Phys. Lett. A},
  volume  = {331},
  number  = {1},
  pages   = {39--44},
  year    = {2004},
  doi     = {10.1016/j.physleta.2004.08.037},
  url     = {https://doi.org/10.1016/j.physleta.2004.08.037},
  eprint  = {quant-ph/0404176},
  archivePrefix = {arXiv}
}

@article{Peschel2003,
  author  = {Peschel, Ingo},
  title   = {Calculation of Reduced Density Matrices from Correlation Functions},
  journal = {J. Phys. A: Math. Gen.},
  volume  = {36},
  number  = {14},
  pages   = {L205--L208},
  year    = {2003},
  doi     = {10.1088/0305-4470/36/14/101},
  url     = {https://doi.org/10.1088/0305-4470/36/14/101}
}

@article{PeschelEisler2009,
  author  = {Peschel, Ingo and Eisler, Viktor},
  title   = {Reduced Density Matrices and Entanglement Entropy in Free Lattice Models},
  journal = {J. Phys. A: Math. Theor.},
  volume  = {42},
  number  = {50},
  pages   = {504003},
  year    = {2009},
  doi     = {10.1088/1751-8113/42/50/504003},
  url     = {https://doi.org/10.1088/1751-8113/42/50/504003},
  eprint  = {0906.1663},
  archivePrefix = {arXiv},
  primaryClass  = {cond-mat.stat-mech}
}

@article{Bose1998,
  author        = {Bose, S. and Vedral, V. and Knight, P. L.},
  title         = {Multiparticle Generalization of Entanglement Swapping},
  journal       = {Phys. Rev. A},
  volume        = {57},
  pages         = {822--829},
  year          = {1998},
  doi           = {10.1103/PhysRevA.57.822},
  eprint        = {quant-ph/9708004},
  archivePrefix = {arXiv}
}

@article{Huhtanen2026,
  author        = {Huhtanen, Santeri and Mafi, Yousef and Moghaddam, Ali G. and Ojanen, Teemu},
  title         = {Many-Body Entanglement Swapping Protocol: Opportunities for Distributed Quantum Computing},
  journal       = {Phys. Rev. Research},
  volume        = {8},
  pages         = {013152},
  year          = {2026},
  doi           = {10.1103/9n1d-7jwj},
  eprint        = {2506.22430},
  archivePrefix = {arXiv},
  primaryClass  = {quant-ph}
}

@article{Lu2022,
  author        = {Lu, Tsung-Cheng and Lessa, Leonardo A. and Kim, Isaac H. and Hsieh, Timothy H.},
  title         = {Measurement as a Shortcut to Long-Range Entangled Quantum Matter},
  journal       = {PRX Quantum},
  volume        = {3},
  pages         = {040337},
  year          = {2022},
  doi           = {10.1103/PRXQuantum.3.040337},
  eprint        = {2206.13527},
  archivePrefix = {arXiv},
  primaryClass  = {cond-mat.str-el}
}

@article{Tantivasadakarn2024,
  author        = {Tantivasadakarn, Nathanan and Thorngren, Ryan and Vishwanath, Ashvin and Verresen, Ruben},
  title         = {Long-Range Entanglement from Measuring Symmetry-Protected Topological Phases},
  journal       = {Phys. Rev. X},
  volume        = {14},
  pages         = {021040},
  year          = {2024},
  doi           = {10.1103/PhysRevX.14.021040},
  eprint        = {2112.01519},
  archivePrefix = {arXiv},
  primaryClass  = {cond-mat.str-el}
}

@article{Cheng2024,
  author        = {Cheng, Zihan and Wen, Rui and Gopalakrishnan, Sarang and Vasseur, Romain and Potter, Andrew C.},
  title         = {Universal Structure of Measurement-Induced Information in Many-Body Ground States},
  journal       = {Phys. Rev. B},
  volume        = {109},
  pages         = {195128},
  year          = {2024},
  doi           = {10.1103/PhysRevB.109.195128},
  eprint        = {2312.11615},
  archivePrefix = {arXiv},
  primaryClass  = {quant-ph}
}

@article{Weinstein2023,
  author        = {Weinstein, Zack and Sajith, Rohith and Altman, Ehud and Garratt, Samuel J.},
  title         = {Nonlocality and Entanglement in Measured Critical Quantum {Ising} Chains},
  journal       = {Phys. Rev. B},
  volume        = {107},
  pages         = {245132},
  year          = {2023},
  doi           = {10.1103/PhysRevB.107.245132},
  eprint        = {2301.08268},
  archivePrefix = {arXiv},
  primaryClass  = {cond-mat.stat-mech}
}

@article{KhannaVasseur2026,
  author        = {Khanna, Kabir and Vasseur, Romain},
  title         = {Measurement-Induced Entanglement in Conformal Field Theory},
  journal       = {Phys. Rev. Lett.},
  volume        = {136},
  pages         = {160402},
  year          = {2026},
  doi           = {10.1103/b7sb-nhjq},
  eprint        = {2508.02788},
  archivePrefix = {arXiv},
  primaryClass  = {quant-ph}
}

@article{Li2018,
  author        = {Li, Yaodong and Chen, Xiao and Fisher, Matthew P. A.},
  title         = {Quantum {Zeno} Effect and the Many-Body Entanglement Transition},
  journal       = {Phys. Rev. B},
  volume        = {98},
  pages         = {205136},
  year          = {2018},
  doi           = {10.1103/PhysRevB.98.205136},
  eprint        = {1808.06134},
  archivePrefix = {arXiv},
  primaryClass  = {quant-ph}
}

@article{Cao2019,
  author        = {Cao, Xiangyu and Tilloy, Antoine and De Luca, Andrea},
  title         = {Entanglement in a Fermion Chain under Continuous Monitoring},
  journal       = {SciPost Phys.},
  volume        = {7},
  pages         = {024},
  year          = {2019},
  doi           = {10.21468/SciPostPhys.7.2.024},
  eprint        = {1804.04638},
  archivePrefix = {arXiv},
  primaryClass  = {cond-mat.stat-mech}
}

@article{Alberton2021,
  author        = {Alberton, Ori and Buchhold, Michael and Diehl, Sebastian},
  title         = {Entanglement Transition in a Monitored Free-Fermion Chain: From Extended Criticality to Area Law},
  journal       = {Phys. Rev. Lett.},
  volume        = {126},
  pages         = {170602},
  year          = {2021},
  doi           = {10.1103/PhysRevLett.126.170602},
  eprint        = {2005.09722},
  archivePrefix = {arXiv},
  primaryClass  = {cond-mat.stat-mech}
}

@article{Poboiko2023,
  author        = {Poboiko, Igor and P{\"o}pperl, Paul and Gornyi, Igor V. and Mirlin, Alexander D.},
  title         = {Theory of Free Fermions under Random Projective Measurements},
  journal       = {Phys. Rev. X},
  volume        = {13},
  pages         = {041046},
  year          = {2023},
  doi           = {10.1103/PhysRevX.13.041046},
  eprint        = {2304.03138},
  archivePrefix = {arXiv},
  primaryClass  = {quant-ph}
}

\appendix
\clearpage

\section{Constructive Gram--Schmidt derivation of the canonical form}
\label{App:CanonicalProof}

\subsection{Adapted occupied-orbital representation}
\label{App:AdaptedChart}

We first choose an occupied-orbital representation adapted to the measured sites.  The existence of this representation for every state with $P_m>0$ is established at the end of Appendix~\ref{App:ProbabilityProof}.  The $m$ measured rows can be completed by $k-m$ rows from $R$ to an invertible $k\times k$ minor.  Denote the selected unmeasured rows by $Q$ and the remaining rows by $T$.  After relabeling,
\begin{align}
L&=\{1,\ldots,m\},
&Q&=\{m+1,\ldots,k\},
\notag\\
T&=\{k+1,\ldots,N\},
&R&=Q\cup T.
\label{Eq:AppendixPartition}
\end{align}
Right multiplication by the inverse of the selected minor gives
\begin{equation}
\widetilde U
=
\begin{pmatrix}
I_m & 0\\
0 & I_{k-m}\\
V_m & Y
\end{pmatrix},
\qquad
V_m=(\mathbf v_1,\ldots,\mathbf v_m).
\label{Eq:AdaptedRepresentative}
\end{equation}
Here $V_m\in\CC^{k\times m}$ and $Y\in\CC^{k\times(k-m)}$.  The columns of $\widetilde U$ form an adapted occupied-orbital basis for the same initial Slater determinant. The first $m$ occupied modes on copy $X=A,B$ are
\begin{equation}
d_{i,X}^\dagger
=
c_{i,X}^\dagger+\widehat v_{i,X}^\dagger,
\qquad
\widehat v_{i,X}^\dagger
=
\sum_{x\in T}(\mathbf v_i)_x c_{x,X}^\dagger.
\label{Eq:AdaptedMeasuredModes}
\end{equation}

\subsection{Single-rung projection and the raw Gaussian state}
\label{App:SingleRung}

For one measured rung, only the pair $d_{i,A}^\dagger d_{i,B}^\dagger$ is affected.  Let
\[
\hat N_i=\hat n_{i,A}+\hat n_{i,B}
\]
be the total particle number on rung $i$. The component in the sector $\hat N_i=1$ is: 
\begin{align}\label{Eq:SingleOccupancyComponent}
\bigl[d_{i,A}^\dagger\wedge d_{i,B}^\dagger\bigr]_{\mathcal N_i=1}
={}&
c_{i,A}^\dagger\wedge\widehat v_{i,B}^\dagger
\notag\\
&-
c_{i,B}^\dagger\wedge\widehat v_{i,A}^\dagger.
\end{align}
Here $\wedge$ denotes the antisymmetrized product of single-particle orbitals; equivalently, in a fixed ordering it is represented by the product of their fermionic creation operators. Using the measured mode in Eq.~\eqref{Eq:MeasuredModeGeneral}, define
\begin{equation}
\eta_i^\dagger(u,v)
=
u^*\widehat v_{i,B}^\dagger-v^*\widehat v_{i,A}^\dagger.
\label{Eq:AppendixResidualRawMode}
\end{equation}
Since
$P_i^{(u,v)}c_{i,A}^\dagger|{\rm vac}\rangle=u^*w_i^\dagger|{\rm vac}\rangle$
and
$P_i^{(u,v)}c_{i,B}^\dagger|{\rm vac}\rangle=v^*w_i^\dagger|{\rm vac}\rangle$,
Eq.~\eqref{Eq:SingleOccupancyComponent} gives
\begin{equation}
P_i^{(u,v)}
\bigl[(d_{i,A}^\dagger\wedge d_{i,B}^\dagger)|{\rm vac}\rangle\bigr]
=
w_i^\dagger(u,v)\wedge\eta_i^\dagger(u,v)|{\rm vac}\rangle.
\label{Eq:GeneralSingleRungProjection}
\end{equation}

Because Eq.~\eqref{Eq:AdaptedRepresentative} associates each measured row with a distinct occupied column, the projector on rung $i$ acts only on the corresponding pair of occupied modes. Iterating Eq.~\eqref{Eq:GeneralSingleRungProjection} therefore preserves the decomposable wedge-product structure.  Ordering row blocks as $(L_A,Q_A,T_A,L_B,Q_B,T_B)$ and column blocks as measured modes, residual modes, $A$ spectators, and $B$ spectators, the raw occupied-mode matrix is
\begin{equation}
W_{\rm raw}^{(u,v)}
=
\begin{pmatrix}
u I_m & 0 & 0 & 0\\
0 & 0 & I_{k-m} & 0\\
0 & -v^*V_m & Y & 0\\
vI_m & 0 & 0 & 0\\
0 & 0 & 0 & I_{k-m}\\
0 & u^*V_m & 0 & Y
\end{pmatrix}.
\label{Eq:RawPostselectedMatrix}
\end{equation}

The first column block contains the fixed measured modes, the second column block contains the generally nonorthogonal residual modes, and the final two column blocks contain the spectator modes in copies $A$ and $B$. Equation~\eqref{Eq:RawPostselectedMatrix} also makes explicit that the conditional state is Gaussian before any normalization or orthogonalization is performed.

\subsection{Gram--Schmidt orthogonalization}
\label{App:GramSchmidt}

The spectator orbitals on $R$ are the columns of
\begin{equation}
U_{\rm sp}
=
\begin{pmatrix}
I_{k-m}\\
Y
\end{pmatrix}.
\label{Eq:SpectatorMatrix}
\end{equation}
Applying Gram--Schmidt to these columns produces an orthonormal set $\{\alpha_\ell\}_{\ell=1}^{k-m}$.  The same invertible column transformation is applied separately to the spectator blocks of copies $A$ and $B$.

Embed each column of $V_m$ into $R=Q\cup T$ by setting its $Q$ components to zero, and continue to denote the embedded vector by $\mathbf v_i$.  The residual vectors are then orthogonalized against the spectator orbitals and against the previously constructed residual vectors:
\begin{subequations}
\label{Eq:GramSchmidtResidual}
\begin{align}
\mathbf b_1
&=
\mathbf v_1
-
\sum_{\ell=1}^{k-m}
\langle\alpha_\ell,\mathbf v_1\rangle\alpha_\ell,
\notag\\[-2pt]
\beta_1
&=
\frac{\mathbf b_1}{\|\mathbf b_1\|},
\label{Eq:GramSchmidtFirst}\\[4pt]
\mathbf b_n
&=
\mathbf v_n
-
\sum_{\ell=1}^{k-m}
\langle\alpha_\ell,\mathbf v_n\rangle\alpha_\ell
\notag\\[-2pt]
&\quad-
\sum_{j=1}^{n-1}
\langle\beta_j,\mathbf v_n\rangle\beta_j,
\notag\\[-2pt]
\beta_n
&=
\frac{\mathbf b_n}{\|\mathbf b_n\|},
\qquad n=2,\ldots,m.
\label{Eq:GramSchmidtLater}
\end{align}
\end{subequations}
The denominators are nonzero whenever $P_m>0$.  Appendix~\ref{App:ProbabilityProof} shows that the Gram matrix of the residual vectors after removal of their spectator components is the positive-definite matrix $\Gamma_m$ in Eq.~\eqref{Eq:GammaChart}.

Every Gram--Schmidt step is implemented by an elementary operation on occupied columns.  If a spectator component $c\alpha_\ell$ is removed from $\mathbf v_i$, one adds $v^*c$ times the $A$-spectator column and $-u^*c$ times the $B$-spectator column to the corresponding residual column.  Likewise, subtracting a previously obtained $\beta_j$ component applies the same spatial subtraction to both copy blocks.  For example,
\begin{equation}
\begin{pmatrix}
-v^*\mathbf v_n\\
u^*\mathbf v_n
\end{pmatrix}
\longmapsto
\begin{pmatrix}
-v^*\bigl(\mathbf v_n-
\langle\beta_j,\mathbf v_n\rangle\beta_j\bigr)\\
u^*\bigl(\mathbf v_n-
\langle\beta_j,\mathbf v_n\rangle\beta_j\bigr)
\end{pmatrix}.
\label{Eq:PairingPreserved}
\end{equation}
Thus the relative coefficients between the two copies are unchanged throughout the orthogonalization.  After normalization, the residual modes are
\begin{equation}
h_j^\dagger(u,v)
=
u^*\beta_{j,B}^\dagger-v^*\beta_{j,A}^\dagger,
\qquad j=1,\ldots,m.
\label{Eq:CanonicalResidualAppendix}
\end{equation}
The resulting orthonormal occupied-mode matrix is
\begin{align}
W_{\rm can}^{(u,v)}
=
[&w_1,\ldots,w_m,
 h_1,\ldots,h_m,
 \alpha_{1,A},\ldots,\alpha_{k-m,A},
\notag\\
&\alpha_{1,B},\ldots,\alpha_{k-m,B}],
\label{Eq:CanonicalMatrixAppendix}
\end{align}
which proves Eqs.~\eqref{Eq:CanonicalStateGeneral} and \eqref{Eq:CanonicalStateFactorization}.

Because the $\alpha_\ell$ and $\beta_j$ orbitals are mutually orthonormal, a fixed fermionic ordering allows the residual state to be written, up to a global phase, as

\[ |\psi_R^{(u,v)}\rangle \doteq |\alpha_A\rangle\otimes|\alpha_B\rangle \otimes \bigotimes_{j=1}^{m} \left( u^*|01\rangle_j-v^*|10\rangle_j \right).
\]

For each $\beta_j$ factor, tracing over its $B$ mode removes the off-diagonal terms because $\langle 0_B|1_B\rangle=0$, leaving $|u|^2 |0\rangle_j\langle0| + |v|^2 |1\rangle_j\langle1|$ on copy $A$. The $B$-spectator state $|\alpha_B\rangle$ is pure and normalized, so its trace gives unity. Therefore, tracing out $B_R$ gives

\begin{equation}\label{Eq:ReducedDensityMatrixAppendix}
\rho_{A_R}= |\boldsymbol\alpha_A\rangle\langle\boldsymbol\alpha_A| \otimes  \bigotimes_{j=1}^{m} \left[ |u|^2|0\rangle_j\langle0| +|v|^2|1\rangle_j\langle1| \right],
\end{equation}
where $|\boldsymbol\alpha_A\rangle=\prod_{\ell=1}^{k-m}\alpha_{\ell,A}^\dagger|{\rm vac}\rangle$.  The first factor is pure, while each $\beta_j$ mode contributes $h_2(|v|^2)$, proving Eq.~\eqref{Eq:VonNeumannEntropyMain}.

\section{Proof of the success-probability formula}
\label{App:ProbabilityProof}

We first prove the formula for states whose measured-row block has full rank, so that the adapted representation in Eq.~\eqref{Eq:AdaptedRepresentative}  can be constructed. Let $|\Psi[\widetilde U]\rangle$ denote the generally unnormalized Slater determinant whose occupied orbitals are the columns of $\widetilde U$, the norm of this doubled initial state is
\begin{equation}
\|\Psi[\widetilde U]_A\Psi[\widetilde U]_B\|^2
=
\det(\widetilde U^\dagger\widetilde U)^2.
\label{Eq:InitialNormAppendix}
\end{equation}
The raw projected state is represented by Eq.~\eqref{Eq:RawPostselectedMatrix}.  With the column order used there, define
$S=I_{k-m}+Y^\dagger Y$ and $K=V_m^\dagger Y$.  Its Gram matrix is
\begin{equation}
G_{\rm post}^{(u,v)} =
\begin{pmatrix}
I_m & 0 & 0 & 0\\
0 & V_m^\dagger V_m & -vK & uK\\
0 & -v^*K^\dagger & S & 0\\
0 & u^*K^\dagger & 0 & S
\end{pmatrix}.
\label{Eq:ProjectedGramMatrix}
\end{equation}
Taking the Schur complement of the two spectator blocks gives
\begin{equation}
\det G_{\rm post}^{(u,v)}
=
\det(S)^2\det\Gamma_m,
\label{Eq:ProjectedNormDeterminant}
\end{equation}
where
\begin{align}
\Gamma_m
&=
V_m^\dagger V_m-KS^{-1}K^\dagger
\notag\\
&=
V_m^\dagger(I_k+YY^\dagger)^{-1}V_m.
\label{Eq:GammaChart}
\end{align}
The coefficients $u$ and $v$ enter the Schur complement only through $|u|^2+|v|^2=1$.

The single-copy Gram matrix is
\begin{equation}
\widetilde U^\dagger\widetilde U
=
\begin{pmatrix}
I_m+V_m^\dagger V_m & V_m^\dagger Y\\
Y^\dagger V_m & S
\end{pmatrix},
\label{Eq:AdaptedGramMatrix}
\end{equation}
and a second Schur complement yields
\begin{equation}
\det(\widetilde U^\dagger\widetilde U)
=
\det(S)\det(I_m+\Gamma_m).
\label{Eq:InitialGramDeterminant}
\end{equation}
The norm ratio therefore gives
\begin{equation}
P_m
=
\frac{\det\Gamma_m}{\det(I_m+\Gamma_m)^2}.
\label{Eq:ProbabilityChart}
\end{equation}
The measured block of the correlation matrix in Eq.~\eqref{Eq:CLDefination} is
\begin{equation}
C_L=(I_m+\Gamma_m)^{-1},
\label{Eq:CorrelationChart}
\end{equation}
so Eq.~\eqref{Eq:ProbabilityChart} becomes Eq.~\eqref{Eq:ProbabilityCorrelation}.

The derivation above used a convenient occupied-orbital basis adapted to the measured region. Such a basis can be chosen whenever $U_L$ has full row rank, so that the projections of the occupied orbitals span all $m$ single-particle modes in $L$. This is the generic situation for $m\leq k$. 

This identity also proves the existence of the adapted representation whenever $P_m>0$.  Indeed, $\det C_L>0$ implies ${\rm rank}\,U_L=m$, while
\begin{equation}
\det(I_m-C_L)
=
\det(I_k-U_L^\dagger U_L)
=
\det(U_R^\dagger U_R)>0
\label{Eq:RankIdentityAppendix}
\end{equation}
implies ${\rm rank}\,U_R=k$.  The rows of $U_R$ therefore span $\CC^k$, and $k-m$ of them can be chosen to complement the $m$ independent rows of $U_L$.  This gives the invertible minor used in Appendix~\ref{App:AdaptedChart}.

Finally, Eq.~\eqref{Eq:RankIdentityAppendix} and $C_L=U_LU_L^\dagger$ give
\begin{equation}
P_m
=
\det(U_L U_L^\dagger)
\det(U_R^\dagger U_R),
\label{Eq:ProbabilityOrbitalAppendix}
\end{equation}
which proves Eq.~\eqref{Eq:ProbabilityOrbitalMatrix}.  Equation~\eqref{Eq:ProjectedNormDeterminant} also establishes directly that the probability is independent of the common normalized pair $(u,v)$.

To connect the probability proof with Gram--Schmidt, embed $V_m$ into $R$ as
\begin{equation}
Z=
\begin{pmatrix}0\\V_m\end{pmatrix},
\qquad
P_{\rm sp}
=
U_{\rm sp}(U_{\rm sp}^\dagger U_{\rm sp})^{-1}U_{\rm sp}^\dagger.
\label{Eq:ResidualProjectionDefinitions}
\end{equation}
A direct substitution gives
\begin{equation}
Z^\dagger(I-P_{\rm sp})Z=\Gamma_m.
\label{Eq:ResidualGramGamma}
\end{equation}
Thus $\Gamma_m$ is the Gram matrix of the residual vectors after their spectator components are removed.  When $P_m>0$, Eq.~\eqref{Eq:ProbabilityChart} implies $\Gamma_m>0$, completing the justification of every denominator in Eq.~\eqref{Eq:GramSchmidtResidual}.

\section{Nonuniform Bell outcomes}
\label{App:NonuniformOutcomes}

To illustrate the role of uniformity, we consider non-uniform Bell post-selection outcomes. We consider a string of Bell outcomes $\boldsymbol{s}=(s_1,\ldots,s_m)$ with $s_i=\pm1$, and define
\begin{equation}
D_{\boldsymbol{s}}
={\rm diag}(s_1,\ldots,s_m).
\label{Eq:MixedSignMatrix}
\end{equation}
A mixed-sign string contains at least one $s_i=+1$ and at least one $s_j=-1$; equivalently, $D_s\neq\pm I_m$. After the spectator components have been removed as in Appendix~\ref{App:GramSchmidt}, let
\begin{equation}
\widetilde V
=
(\widetilde{\mathbf v}_1,\ldots,\widetilde{\mathbf v}_m),
\qquad
\Gamma=\widetilde V^\dagger\widetilde V>0.
\label{Eq:MixedProjectedMatrix}
\end{equation}
The residual occupied-mode block is
\begin{equation}
H_{\rm raw}^{(\boldsymbol{s})}
=
\frac{1}{\sqrt2}
\begin{pmatrix}
-\widetilde V D_{\boldsymbol{s}}\\
\widetilde V
\end{pmatrix},
\label{Eq:MixedRawBlock}
\end{equation}
with Gram matrix
\begin{equation}\label{Eq:MixedGramMatrix}
\bigl(H_{\rm raw}^{(\boldsymbol{s})}\bigr)^\dagger
H_{\rm raw}^{(\boldsymbol{s})}
= \frac12 \left( \Gamma+D_{\boldsymbol{s}}\Gamma D_{\boldsymbol{s}} \right).
\end{equation}

Writing $H_{\rm raw}^{(s)}=(H_A^{\mathsf T},H_B^{\mathsf T})^{\mathsf T}$, the residual correlation matrix on $A_R$ is
\[
C_{A_R}^{\rm res}
=
H_A
\bigl[
(H_{\rm raw}^{(s)})^\dagger H_{\rm raw}^{(s)}
\bigr]^{-1}
H_A^\dagger .
\]
The nonzero eigenvalues  of the correlation matrix restricted to $A_R$ in this residual sector are the eigenvalues of
\begin{equation}
\bigl(D_{\boldsymbol{s}}\Gamma D_{\boldsymbol{s}}\bigr)^{1/2}
\left(
\Gamma+D_{\boldsymbol{s}}\Gamma D_{\boldsymbol{s}}
\right)^{-1}
\bigl(D_{\boldsymbol{s}}\Gamma D_{\boldsymbol{s}}\bigr)^{1/2}.
\label{Eq:MixedCorrelationEigenvalues}
\end{equation}

The residual contribution to the von Neumann entropy is therefore
\begin{equation}
S_{A_R}^{\rm res}
=
\sum_{q=1}^{m}h_2(\lambda_q)
\leq
m\ln2.
\label{Eq:MixedEntropyBound}
\end{equation}
Equality holds if and only if every $\lambda_q=1/2$, which is equivalent to
\begin{equation}
D_{\boldsymbol{s}}\Gamma D_{\boldsymbol{s}}
=
\Gamma.
\label{Eq:MixedMaximalityCondition}
\end{equation}
For a genuine mixed string, this requires $\Gamma_{ij}=0$ whenever $s_i\neq s_j$.  It is a special block-orthogonality condition on the initial-state-dependent residual orbitals and is not satisfied generically.

For $m=2$, take $D_{\boldsymbol{s}}={\rm diag}(1,-1)$ and $\Gamma=\left(\begin{smallmatrix}1&r\\r^*&1\end{smallmatrix}\right)$.  Equation~\eqref{Eq:MixedCorrelationEigenvalues} gives $\lambda_\pm=(1\pm|r|)/2$.  Unless $r=0$, Eq.~\eqref{Eq:MixedEntropyBound} is strict.  Here $r$ is the off-diagonal Gram-matrix element between the spectator-projected residual vectors in the two sign sectors. Thus, maximality and equivalently the universal $m$-Bell-pair form of the uniform branch requires all cross-sign Gram-matrix elements to vanish, in agreement with Eq.~\eqref{Eq:MixedMaximalityCondition}.

\end{document}